\documentclass[12pt]{iopart}
\usepackage{hyperref}
\usepackage{amsfonts}

\begin{document}

\title[A note on vacuum energy from the de Sitter spectrum]{A note on vacuum energy from the de Sitter spectrum}
\author{Kostyantyn Ropotenko}
\address{State Department of communications and
informatization, Ministry of transport and communications of
Ukraine, 22, Khreschatyk, 01001, Kyiv, Ukraine}

\begin{abstract}
It is shown that a well-known relation between entropy of a system
and its energy spectrum being applied to the early universe
determines the present vacuum energy and the time scale on which
this energy can manifest itself. Given the present vacuum energy,
the relation imposes a constraint on the initial inflationary state.
\end{abstract}


\bigskip\bigskip

There is strong evidence in favor of an accelerating universe which
might be due to a tiny cosmological constant \cite{Ast}. This means
that the universe will asymptotically tend to the de Sitter state.
In accordance with the inflationary cosmology, such a state existed
also in the distant past, although with much more vacuum energy. In
this connection, an important question arises: is there a relation
between these states? Even if it is the case, then why does the
present vacuum energy density have the value it has, $\rho
\sim10^{-47}~{\rm GeV}^{4}$? Where does such a small number come
from? It is so small in comparison with the expected Planckian scale
$\rho_P\sim10^{76}~{\rm GeV}^{4}$ that its relation with the early
inflationary epoch is considered to be unlikely \cite{Bousso}.

In this note, I want to point out to a possible relation between the
energy spectrum of de Sitter space in the early universe and the
present vacuum energy density.

My proposal rests on the conception of entropy. As is well known,
the entropy of a system $S$, by definition, is the logarithm of the
number of states with energy between $E$ and $E+\delta{E}$. The
width $\delta{E}$ is some energy interval characteristic of the
limitation in our ability to specify absolutely precisely the energy
of a macroscopic system. It is equal in order of magnitude to the
mean fluctuation of energy of the system. Dividing $\delta{E}$ by
the statistical weight $e^{S(E)}$, we obtain the mean separation
between levels in the interval \cite{Land}
\begin{equation}
\label{main}\Delta{E}=\delta{E}\: e^{-S(E)}.
\end{equation}

This expression can be immediately applied to de Sitter space. As is
well known, it is a thermodynamical system with a temperature and an
entropy given by
\begin{equation}
\label{temp}T=\frac{H}{2\pi}
\end{equation}
\vskip .5cm
\begin{equation}
\label{ent}S=\frac{\hspace{0cm}\mbox{\emph{Horizon
area}}}{4G}=\frac{\pi M_{P}^{2}}{H^{2}}\,,
\end{equation}
where $H$ is the Hubble constant. Taking into account the first law
of thermodynamics of de Sitter space $dE=TdS$ \cite{GH,SSV},
expressed in terms (\ref{temp}) and (\ref{ent}), we can also define
the specific heat
\begin{equation}
\label{heat}C_{v}=\left(\frac{\partial{E}}
{\partial{T}}\right)_{V}=\frac{1}{2\pi\, G\,T^{2}}=\frac{2 \pi
M_{P}^{2}}{H^{2}}.
\end{equation}

Then, since the the mean square fluctuation of energy is
$\langle(\delta{E})^{2}\rangle=C_{v}T^{2}$, it follows that
\begin{equation}
\label{rsm}\delta{E}\sim M_{P}.
\end{equation}

Thus
\begin{equation}
\label{fin}\Delta{E}\sim M_{P}\:e^{-S(E)}.
\end{equation}

In accordance with Bohr's frequency condition, (\ref{fin}) should
equal to some frequency. The only frequency in the problem is the
rate of expansion, i.e. the Hubble constant $H$. Such a choice also
agrees with the uncertainty principle $\Delta{E}~\Delta{t}\sim 1$,
where $\Delta{t}\sim H^{-1}$. Thus we obtain

\begin{equation}
\label{final}M_{P}\:e^{-S(E)}\sim H.
\end{equation}

Notice that the value of $H$ on the right-hand side of (\ref{final})
is, of course, not the same as that in the argument of the entropy
$S(E)$ on the left-hand side. The value of $H$ on the left-hand side
is given as initial one. The value of $H$ on the right-hand side, in
contrast, must be found; it correspondents to a space whose rate of
expansion is the same as the mean energy spacing (\ref{final}). In
order to distinguish these values, we shall write them with
different indexes: $H_{i}$ for a given (initial) value and $H_{f}$
for a required (final) one.

Let us estimate $H_{f}$. Consider the early universe. Reasonable
scale of inflation ranges from the Planck scale $M_{P}\sim
10^{19}~{\rm GeV}$ ($H_{i}\sim 10^{19}~{\rm GeV})$ to the GUT scale
$M_{GUT}\sim10^{16}~{\rm GeV}$ ($H_{i}\sim 10^{13}~{\rm GeV})$. In
accordance with (\ref{ent}), it correspondents to the range of
$S_{i}\sim10^{0}-10^{12}~(H_{i}\sim 10^{19}-10^{13}~{\rm GeV})$. Now
if we take $S_{i}\sim10^{2.15}~(H_{i}\sim 10^{17.8}~{\rm GeV})$ from
the range and use (\ref{final}), we obtain $H_{f}\sim10^{-42}~\rm
GeV$ corresponding $\rho_{f}\sim 10^{-47}~{\rm GeV}^{4}$
($H=\sqrt{\frac{8\pi G \rho}{3}}$). On the other hand, the
finiteness of the de Sitter entropy indicates that spectrum of
energy is discrete. The discreteness of the spectrum means that
there is a typical energy spacing (\ref{final}). It defines a new
time scale $t_{f}\sim H_{f}^{-1}$ of order $\sim 10^{17}~\rm s$.
This is the time scale on which the discreteness of the spectrum can
only manifest itself. Therefore, it is natural to identify $H_{f}$
and $\rho_{f}$ with the present Hubble constant and the vacuum
energy density.

We can also identify the time $t_{f}$ with the Poincar\'{e}
recurrence time. The quantum Poincar\'{e} Recurrence theorem states
\cite{DKS}: given a system in which all energy eigenvalues are
discrete, a state will return arbitrarily close to its initial value
in a finite amount of time. These Poincar\'{e} recurrences generally
occur on a time scale exponentially large in the thermal entropy of
the system. Thus we define the Poincar\'{e} recurrence time
$t_{r}\equiv t_{f}=M_{P}^{-1}e^{S(E)}$. We can say that the universe
returns to its initial point to within the mean energy spacing
$\Delta{E}\sim 10^{-42}~\rm GeV$ in the Poincar\'{e} recurrence time
$t_{r}\sim 10^{17}~\rm s$; this process has became noticeable with
the detection of the cosmic acceleration.

We can also apply (\ref{final}) to the present universe. Using the
current value of the entropy $S_{i}\sim 10^{120}$ we obtain
$H_{f}\sim 10^{-119}~\rm GeV$. But this gives us nothing
interesting.

In this note we apply a well-known thermodynamical relation between
entropy of a system and its energy spectrum (\ref{main}) to the
early universe. It turns out that, despite the absence of a theory
of quantum gravity, one can relate the de Sitter state of the early
universe with that of today's universe and determine the present
vacuum energy and the time scale on which this energy can manifest
itself. Moreover, as is seen from (\ref{final}), such a transition
from the inflationary vacuum energy to the present value is related
to a quantum transition between the energy levels of the initial
state. On the other hand, the relation imposes a constraint on the
initial inflationary state. To get a universe with the vacuum energy
density $\rho\sim10^{-47}~{\rm GeV}^{4}$ today, we should have the
universe at the beginning with the following parameters.

\begin{itemize}
  \item Hubble constant $H_{i}\sim 10^{17.8}~\rm GeV$
  \item Entropy $S_{i}\sim10^{2.15}$
  \item Energy $E_{i}\sim 10^{20.2}~\rm GeV$
  \item Width of the spectrum $\delta{E}\sim10^{19}~\rm GeV$
  \item Mean energy spacing $\Delta{E}\sim10^{-42}~\rm GeV.$
\end{itemize}

That is, instead of asking why does the present vacuum energy
density have the value $\rho\sim 10^{-47}~{\rm GeV}^{4}$, we can ask
why did the universe have the value $\rho_{i}\sim 10^{73.6}~{\rm
GeV}^{4}$ at the beginning.

\section*{References}
\bibliographystyle{iopart-num}

\end{document}